# ObjectEMS: Electrical Muscle Stimulation Without Electrodes on the User

Yun Ho

University of Chicago, Chicago, Illinois, USA, yunho@uchicago.edu

Zhechen Zhao

University of Chicago, Chicago, Illinois, USA, zhechen@uchicago.edu

Romain Nith

University of Chicago, Chicago, Illinois, USA, rnith@uchicago.edu

Andre de la Cruz

University of Chicago, Chicago, Illinois, USA, andredlcruz@uchicago.edu

Pedro Lopes

University of Chicago, Chicago, Illinois, USA, pedrolopes@uchicago.edu

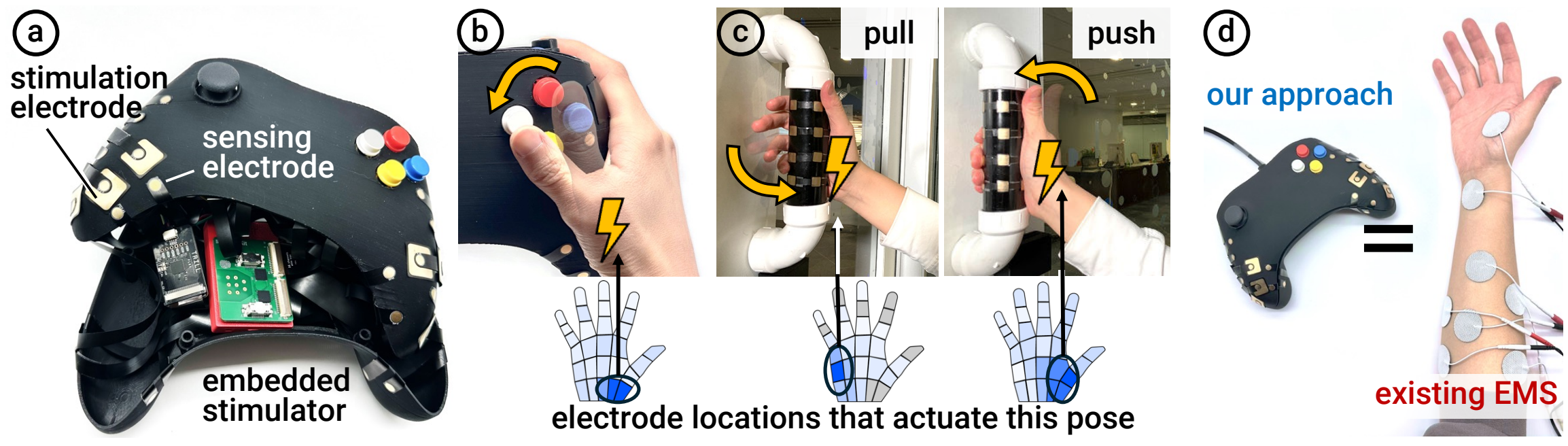

Figure 1: To assist with interactive tasks, electrical muscle stimulation (EMS) systems demand that users attach electrodes and stimulators to their bodies. To explore an alternative route, we invert this traditional approach and place all electrodes (and stimulator) inside objects that the user interacts with: (a) we developed self-contained proof-of-concept objects with embedded electrodes and a stimulator inside the object, such as this game controller with force feedback. Through our empirical validation, focused on the palm (i.e., the body surface that most commonly makes contact with objects), we found the electrode locations for finger actuations from the palm. This informed strategic locations of electrodes on this exemplary (b) controller and (c) responsive door handles that prevent push-pull confusion. (d) By moving the electrodes to objects, our approach frees up the user's body.

**Abstract**. Interactive electrical muscle stimulation (EMS) has revealed its promise as a portable interface for force-feedback. However, while much ink has been spilled about the advantages of EMS, few have investigated one of its central limitations: the need to attach electrodes to users. This has dramatically limited the application of EMS—especially in brief interactions or physical assistance with tools. To explore an alternative, we propose embedding electrodes (and stimulator) inside objects that the user interacts with. This is made possible because we identified multiple novel electrode placements that can elicit four distinct finger movements from the palm (no forearm stimulation). To illustrate this new way of implementing electrical muscle stimulation, we developed a set of self-contained interactive objects that use capacitive sensing to determine if a user's hand is in poses conducive to stimulation and then actuate the fingers from contact points between the hand and the grasped object. In our user study, we found that participants spent less time calibrating *ObjectEMS* than traditional EMS, and they felt less tethered while engaging with an EMS application via this novel approach. Our approach frees up the user's body from wearing electrodes and EMS stimulators, providing a new pathway for researching, implementing, or deploying actuated tangible interfaces. We demonstrate its potential via exemplary applications, such as a game controller with finger-level force feedback, a door handle with force feedback to prevent push-pull confusion, and more.

## 1 INTRODUCTION

Over a decade of research in interactive electrical muscle stimulation (EMS) has revealed its promise as a portable interface to achieve force feedback without large mechanical actuators [49]. For example, *Affordance++* [54] and *generative muscle stimulation* [28] provide physical assistance to operate everyday objects (e.g., using a spray can, a camera, etc.). However, to benefit from these systems, users need to wear electrodes on the correct locations and wear hardware stimulators. This points to a key limitation of EMS: the *burden* of wearing hardware and electrodes ahead of the desired interaction *is on the user* [83].

The time required to attach electrodes to the user's body runs contrary to most application areas in which EMS systems have been researched, i.e., brief and casual everyday interactions, such as rolling a bowling ball [84] or taking a photo [37]. In fact, much of contemporary HCI research has emphasized this as a key limitation of EMS. For instance: wearing the electrodes "felt more like a medical device" [80]; "It is difficult to fasten the [electrodes]" [82]; and "the placement of the electrodes (...) is challenging" [68].

Many have explored improving the form factor of current wearable EMS systems, such as by reducing the hardware footprint to fit inside other devices (e.g., smartwatch [80] or mobile phone [51]) or specialized garments (e.g., sleeves [14, 41], suits [87], and so forth). Unfortunately, these approaches still result in devices that users are tasked with wearing and, therefore, inherit the aforementioned issues of body-worn devices.

We explore the opposite concept, which we call *ObjectEMS*. Rather than having electrodes on the user's body, we attach electrodes to objects that the user interacts with, such as a game controller with finger-level EMS force feedback in Figure 1 (b) or (c) door handles that, despite their push-pull affordance issues (classic "Norman Doors" [61]), use EMS force-feedback to make the user's hand perform the push-pull gestures (pull via pinky flexion, push via thumb flexion).

Our novel idea of reversing the primary way that EMS electrodes are deployed is not without technical challenges. First, we need to validate the feasibility of actuating *from the palm* (i.e., the surface of the body that most typically makes contact with objects [5, 23]). We conducted an empirical validation and identified four distinct actuated movements from the vantage point of the hand (i.e., thumb flexion, thumb abduction, pinky flexion, thumb and pinky flexion). We leverage this knowledge to design the EMS electrode layouts and actuations via *ObjectEMS*.

However, users can be in contact with the electrodes in many ways, including hand poses conducive to stimulation (i.e., likely to generate movement) and unsuitable poses (i.e., likely to generate only tactile sensations). To tackle this, our system estimates the hand poses via capacitive touch (with an accuracy of ~74.10% in our technical experiment) and only stimulates when the system detects that the hand is in a conducive pose for actuation. While there are many technical routes to implement our concept (e.g., many approaches to pose tracking), our key contribution is not to exhaustively implement all of these sensing alternatives, but to demonstrate the concept of moving the EMS electrodes (and stimulator) to *objects* that the user interacts with.

To contrast our approach with traditional body-worn EMS, we conducted a user study that required participants to calibrate EMS by themselves and to operate an EMS application while also engaging in social interactions with others. We observed that participants found it easier to calibrate *ObjectEMS*, and they spent less time calibrating, than in traditional EMS. They also found it easy to switch between *ObjectEMS* and other situations (e.g., social interactions); comparatively to this, they felt their mobility was impaired by the cable length of the traditional EMS approach.

Finally, we believe this approach provides a new pathway to researching, implementing, or even deploying interactive EMS systems. While it is not without limitations, it is well-suited for EMS applications that involve brief interactions, such as EMS assistance with tangible & everyday objects. We demonstrate applications that arise from the benefits of our approach, such as the aforementioned game controller, an actuated guitar, an interactive workout dumbbell, etc.

## 2 RELATED WORK

Our work primarily builds on interactive electrical muscle stimulation (EMS). We narrow in on the limitations of EMS, namely, its dependence on body-worn electrodes. Finally, since our approach is based on moving the electrodes to objects, we briefly discuss and draw inspiration from works using sensing/actuation in tangibles.

### 2.1 Interactive Electrical Muscle Stimulation (EMS)

EMS' origins can be traced to medicine/neuroscience, as a way of rehabilitating muscle [2, 10, 20, 22, 29, 77, 78]. Two decades ago, it emerged in HCI as a popular approach to providing force feedback by electrically actuating the user's muscles [44, 82]. The popularity of interactive EMS stems from its small form factor relative to mechanical-based force-feedback devices, such as exoskeletons or other mechanical actuators [50]. These benefits led to ~150 EMS publications in HCI [18], spanning applications such as AR/VR [40, 53, 55], or skill-acquisition [17, 26, 54, 84].

### 2.2 Critical limitations of EMS

Despite this, there are still several limitations that hinder EMS from wider applications, many of which researchers have addressed in recent years. Output accuracy has been improved through novel electrode placements [56, 79, 82] or mechanical designs [60]. Accuracy has been further refined through closed-loop control [11, 38, 56], biomechanical models [30, 31], or even signal-level techniques [62, 80]. Calibration overhead has also been improved by user-controlled parameters [69, 80] or automation [42, 90]. Considerable effort has also gone into improving wearability through sleeves [41] and suits [87], all to make wearing EMS electrodes more practical. Yet, all of these systems retain one fundamental constraint: *electrodes must still be placed directly on the user's skin.*

### 2.3 A direct attempt to forego electrodes (magnetic coils)

The most direct attempt to address this replaces electrical with magnetic coils for muscle stimulation [83], foregoing skin-contact electrodes entirely. Unfortunately, this requires expensive and, more importantly, extremely *large* electromagnetic coils and stimulators, thus making it only useful for scenarios in which very large objects can be embedded with such heavy instrumentation (the coils used in this approach are 30x larger than a typical electrode, and the stimulator is ~1350x larger than an EMS stimulator [67]). Despite advances to reduce the size of magnetic stimulators, this approach is inherently limited by electromagnetic laws, i.e., to produce an extremely strong magnetic field, large currents are needed (thus, requiring large capacitor bands, large coils, etc.) As such, even the most cutting-edge stimulators are the size of an entire backpack, at 11×27×29 cm and weighing 9 kg [9]. By contrast, EMS stimulator [67] (4×6×1.7 cm) is ~210x smaller, not to mention the negligible size and weight of EMS electrodes. Finally, magnetic stimulation generates up to 80.7 dB of loud acoustic noise [83], necessitating ear protection.

### 2.4 Sensing and actuation from tangible objects

Much research has targeted integrating input (sensing) and output (actuation) into interactive tangible objects. We provide only a succinct review—refer to [16, 75] for extensive reviews.

#### *2.4.1 Hand pose sensing for tangibles*

A canonical way to enable touch sensing on objects is capacitive sensing, which is already used in commercial VR controllers [88]. Many works [7, 25, 64, 70, 73, 74, 92] have embedded capacitive electrodes in objects to detect touch location, contact area, or poses. Additionally, sweep-capacitive [72] or even electric field tomography [98] can detect complex poses or support irregularly shaped objects. More recently, entire fabrication pipelines have been developed for retrofitting 3D objects with real-time multi-touch sensing without modifying their interiors [32].

While we draw direct inspiration from capacitive (given its long-standing use for hand pose sensing), many other alternative methods exist, including acoustics [46, 63, 85], optics (fiber optics [4, 6, 94] or gel-based optical methods [96]), and more. Our focus is not to innovate in sensing nor to exhaustively compare these sensing alternatives, but to focus on moving the EMS electrodes to the surface of objects that the user interacts with. Thus, we draw inspiration from tangible actuators.

#### *2.4.2 Force feedback (actuation) on tangibles*

While vibration is one of the most dominant haptic actuators found in tangibles, this modality lacks the force to physically displace the user's hand. As such, researchers have turned to a wide variety of actuators or even illusions to endow tangibles with hardware to actuate (i.e., displace) the body.

One way that researchers have actuated the body from tangibles is via illusions, such as asymmetric vibrations that induce a pulling sensation (without physical displacement) [39, 71]. To render stronger kinesthetic effects, many have embedded spinning motors inside tangibles to exploit gyroscopic forces [3, 57, 95], or motorized actuators of various kinds (e.g., servos, pneumatics, hydraulics) that can physically push against the user [12, 13, 19, 21, 27, 36, 43, 58, 76, 86, 91, 93, 97]. While we draw inspiration from these works, they tend to exhibit the same well-known limitation: it is strikingly hard to miniaturize sufficiently strong motors to actuate body parts [49]—this leads to tangibles that can move the body, but rarely can pack all the actuators, power source, and necessary hardware *inside* an actual object. In contrast, many have argued that EMS provides an extremely compact form of force feedback, which we leverage in our approach to actuating the body from electrodes on tangibles.

### 2.4.3 Electrical (tactile, not muscle) actuation on tangibles

Finally, an approach from which we draw significant inspiration is *Tactlets* [24], which also embeds electrodes in objects but uses them to deliver electro*tactile* feedback when a user contacts the object: i.e., the user feels touch (no movement). Electrotactile operates on a fundamentally different mechanism than EMS: (1) it stimulates cutaneous nerve endings near the skin surface, thus arising sensations of touch and does not move muscle fibers (which are deeper in the body); (2) usually, nearby electrodes (as close as a few mm or less) are used to create a sensation near the electrodes—in fact, our empirical results, precisely show that EMS works differently, by requiring a significantly larger separation of electrodes (even for the case of a thin body part like the hand) to induce a deeper current path to successfully recruit muscles [34, 62, 80]; (3) due to the latter phenomena (deeper current paths requiring longer electrode distance), increasing the intensity of electrotactile stimulation does not always induce a physical movement; and (4) while it was well-known that electrodes on the palm cause skin sensations via electrotactile [33, 34, 47, 66, 67, 81], it is a novel contribution of our work to show that electrodes on the palm can induce up to four distinct finger movements.

## 3 WALKTHROUGH

To illustrate the applicability of our concept, we detail a walkthrough akin to our second user study, where a user is playing *Mario Kart* with force-feedback via EMS—force-feedback is one of the most active research areas of EMS in HCI [17, 35, 38, 51]. This walkthrough is meant to contrast the traditional approach (i.e., EMS electrodes worn on the body) with our approach, which moves electrodes and EMS hardware to objects that the user interacts with. In Figure 2, we see the user starting their console gaming experience with a game controller: (a) In current EMS systems, they are forced to attach electrodes and stimulators ahead of time; in contrast, (b) our approach moves electrodes to the game controller, so the user feels less tethered than with traditional EMS electrodes and stimulators. The user can, for instance, more rapidly start the gaming session to enjoy force feedback. In fact, in our second user study (where participants played a similar game of *Mario Kart* as depicted here), we found that, indeed, participants experienced less friction in switching back to the game with *ObjectEMS* after being interrupted, as they could hold the controller again to resume the game without the need for re-attaching the electrodes to their bodies.

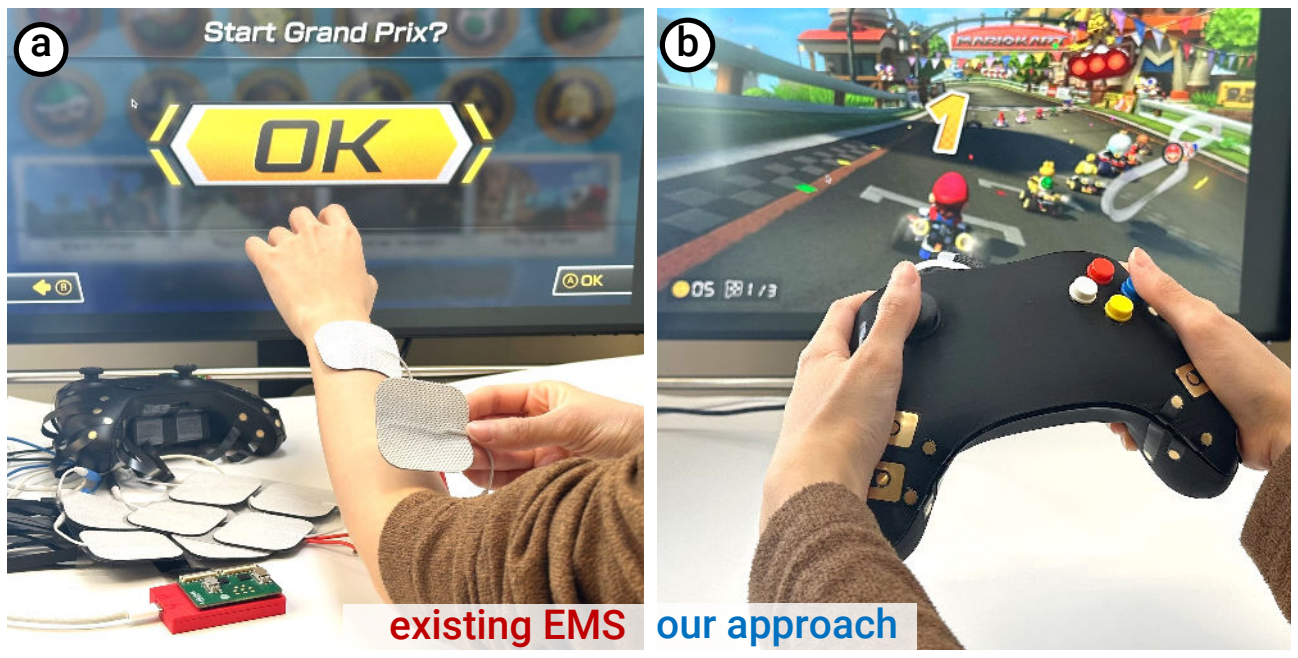

Figure 2: (a) With EMS systems, users must attach electrodes and stimulators ahead of time—here, before starting the game. (b) With our approach (i.e., moving the electrodes to the game controller), the user more rapidly starts their gaming session.

As depicted in Figure 3, in this gaming experience, EMS provides force feedback during in-game events: (a) rather than using traditional forearm-based stimulation to press down the user's finger when they collect a "speed boost" powerup (thus, pressing their finger involuntarily to accelerate); our approach (b) achieves thumb actuation from only the vantage point of where the palmar side of the hand grasps the controller—as illustrated in the "zoomed-in" view of Figure 3 (b). Note that our approach achieves this without the need for any electrode below the hand, which is typically needed for thumb actuation in existing EMS systems (e.g., [54, 59, 79]).

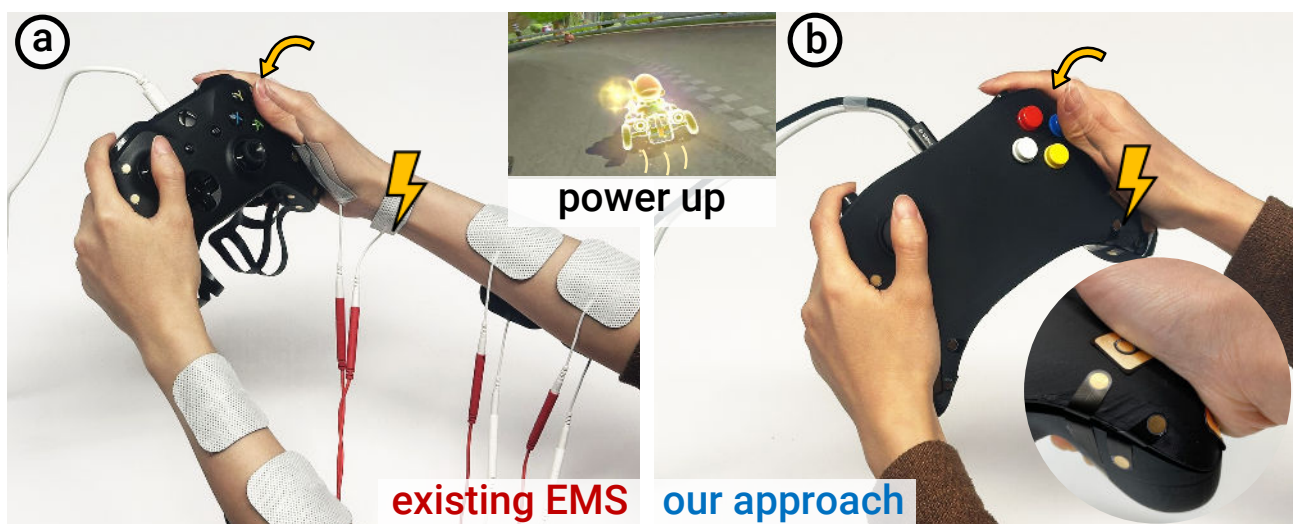

Figure 3: An actuated push button (via thumb flexion) with (a) an existing EMS system and (b) our approach. Note the zoom (bottom right), showing the electrodes under the right palm.

Next, due to the previous speed boost, the user needs to drift around a curve. Again, EMS provides a rumble-type force feedback, as depicted in Figure 4: (a) via existing systems, actuation of both hands moving left and right to achieve rumble requires attaching electrodes to both forearm flexors (e.g., as in [45, 51]); instead, our approach (b) actuates the finger flexors to also cause rumble when hands contact electrodes on the controller that are at strategic locations.

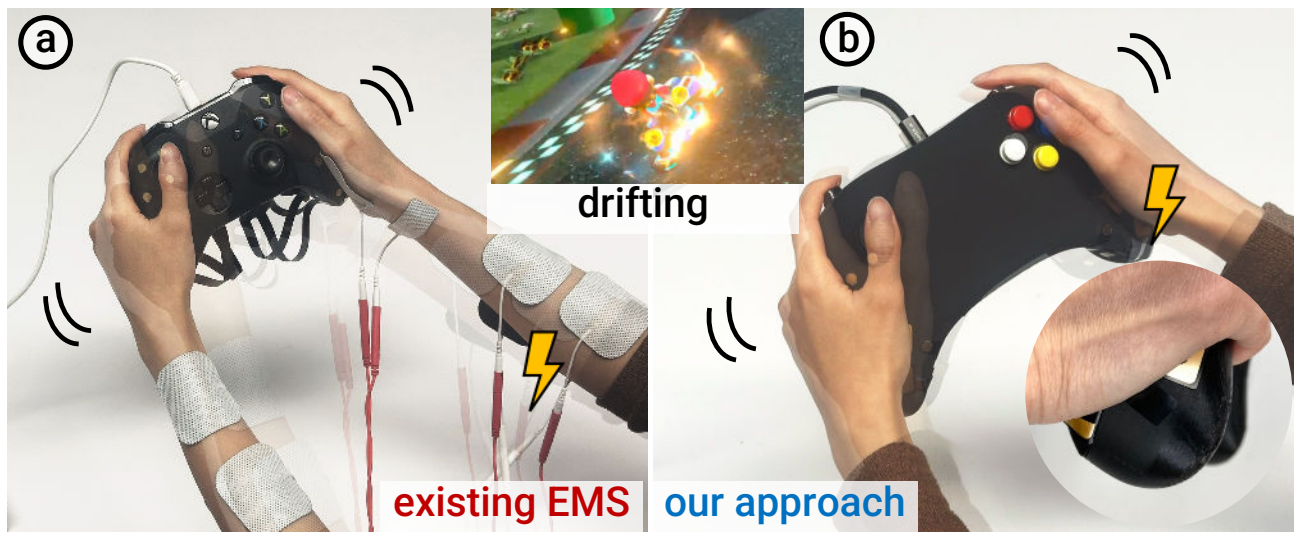

Figure 4: Actuated rumble (via pinky flexion on both hands) with (a) an existing EMS system and (b) our approach. Note the zoom (bottom right) shows the electrodes under the palm.

Now, the opponent blocked their ability to activate power-ups by deploying a "lightning strike" attack. EMS renders this as force feedback by preventing the user from pressing buttons while the attack is active: (a) existing EMS systems that prevent button presses (e.g., [52]) achieve this by stimulating the extensors at the dorsal side of the forearm; instead, (b) our approach stimulates index extension merely from palmar locations that rest on the controller's surface.

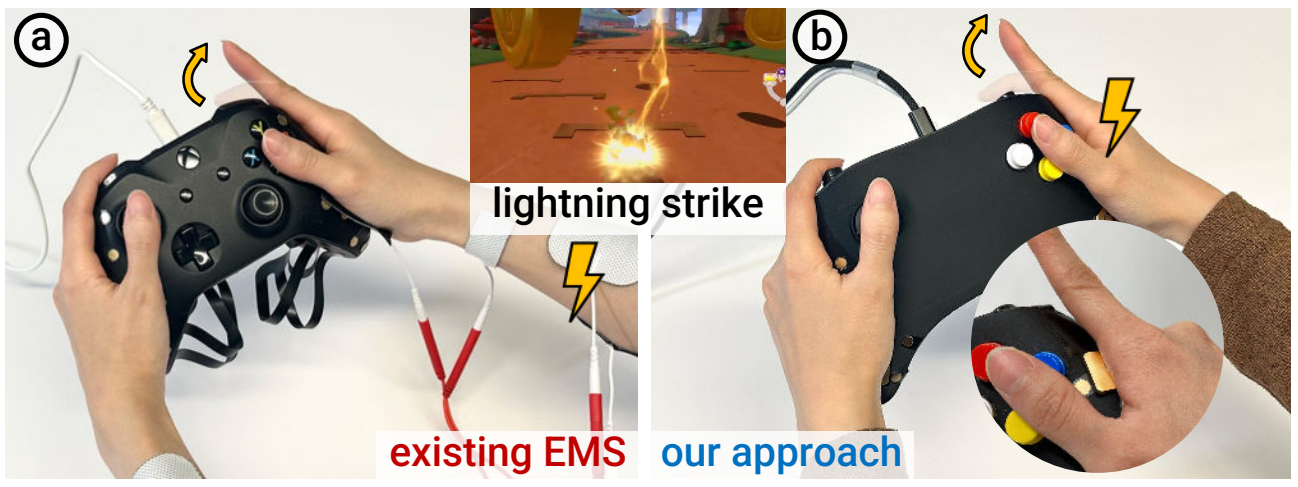

Figure 5: Preventing the user from pressing buttons (via index extension) with (a) an existing EMS system and (b) our approach. Note the zoom (bottom right) shows the electrodes under the finger.

In the final stretch before the finish line, our user gets a power-up that gives them speed and automatic steering. EMS renders this as force feedback by automatically steering in this final curve while the power-up is active. Our approach can realize this left thumb abduction merely from contact points between the hand and the controller's surface.

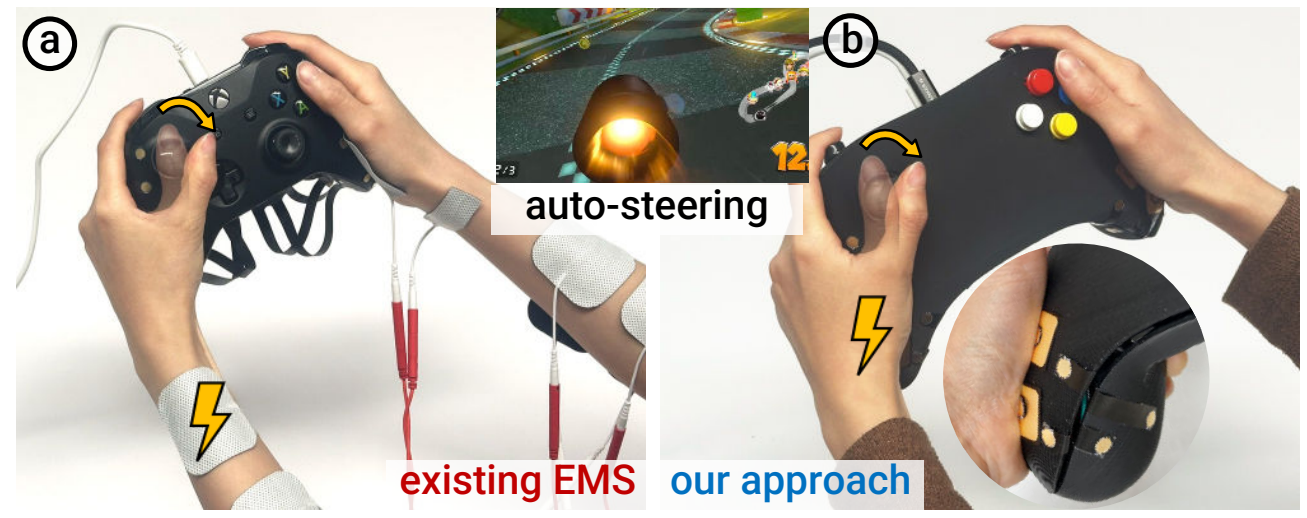

Figure 6: Moving the joystick (via thumb abduction) with (a) an existing EMS system and (b) our approach. Note the zoom (bottom right) shows the electrodes under the left palm.

Finally, while this walkthrough is not meant to explicate our hardware design (see *Proof of concept* for details), we succinctly illustrate the built-in sensing that enables this game controller to sense the user's hand poses. In addition to stimulating electrodes, our objects also feature capacitive sensing electrodes, which are monitored to ensure the user's hands are in poses conducive to stimulation (which we found to achieved 74.10% accuracy). Figure 7 depicts two examples: (a) this typical "dual-handed grasp" pose is conducive to four possible actuations; whereas (b) this "holding the controller on the bottom right" is not detected as a pose that can trigger any of our actuations.

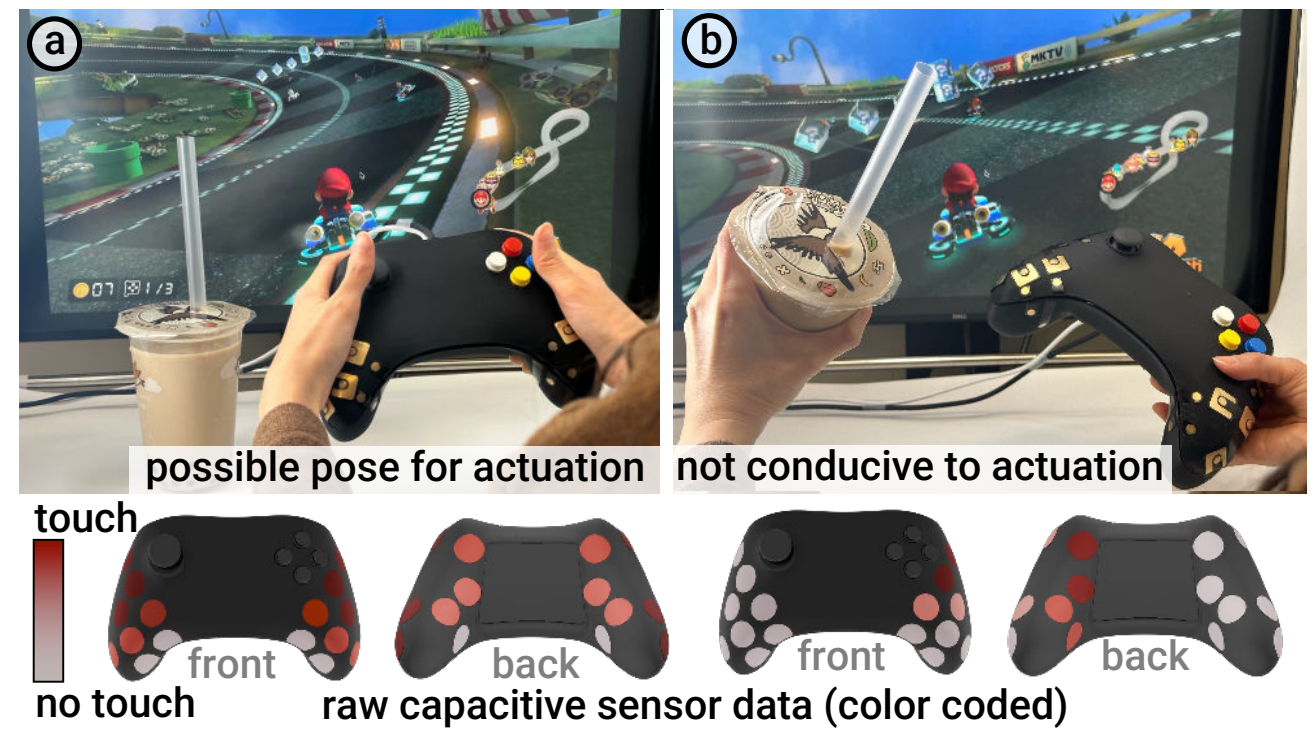

Figure 7: *ObjectEMS* features capacitive sensing electrodes to test if the user's hands are in poses conducive to stimulation: (a) this capacitive touch profile was detected by the system as conducive to actuation. As such, if the game renders any force feedback event, stimulation will proceed. Conversely, (b) this capacitive touch profile was not detected by the system to match any of its actuation poses; no EMS will be delivered.

## 4 EMPIRICAL VALIDATION: IS IT POSSIBLE TO MOVE THE HAND WITH EMS... FROM THE HAND?

Our novel idea of moving the attachment location of EMS electrodes (i.e., from the body to objects) is not without a technical challenge requiring empirical validation to be overcome: *Can we actuate hand poses by means of EMS when the electrodes are placed on the hand itself?* Thus, in this experiment, we stimulated—with EMS—the most detailed map of the palm to date and observed (using cameras) any resulting finger movements. This was approved by our ethics board (IRB19-1949).

### 4.1 Calculating our 378 stimulation targets on the palm

We turned our attention to the palmar side of the hand, since it is the most common way that grasp-based interactions take place [5, 23, 52, 65]. We followed the anatomical structure of the palm to calculate our stimulation targets. First, each finger was divided into its sections, resulting in three areas per finger (one per phalange) and two for the thumb (one per phalange). Then, this tripartite subdivision of the finger was propagated to the palm, anatomically following the tendon that connects the finger to the wrist, yielding three more locations per finger in the palm, and two for the thumb. In total, this resulted in 28 segments of the palm, as depicted in Figure 8 (a). Since EMS requires a current return path from the stimulation point, two electrodes are required. This yields $C(28,2) = 378$ positions via mathematical combinations.

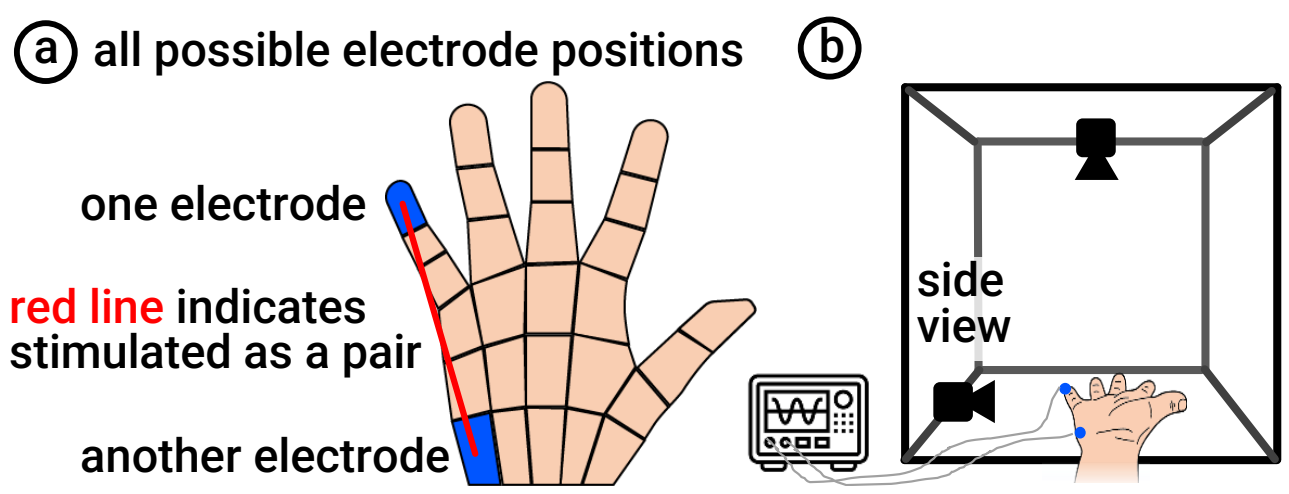

Figure 8: (a) Grid of 378 electrode positions in our empirical validation. (b) Setup of the cameras and stimulator.

### 4.2 Apparatus

Figure 8 (b) shows the setup: two cameras filmed the dominant hand (side and top views) to capture involuntary (stimulated) movements. Camera images were corrected with the checkerboard method for any lens distortion. Stimulation was delivered using 10 × 10 mm electrodes via a portable medical-grade stimulator (*P24* from *HASOMED*).

### 4.3 Experimental procedure

We recruited six participants (three female and three male; 26 ± 2.8 years old; all right-handed), compensated monetarily for their time. The study took up to seven hours per subject for a comprehensive exploration of hand stimulation (i.e., 378 combinations of electrode positions). Each participant was stimulated at a total of 378 electrode positions, which brings our total to 2,268 trials across participants—likely the most comprehensive EMS study known to date on the hand.

Per trial, we placed two electrodes at two randomly chosen grid positions (one pair in 378 possible paired combinations). Stimulation was increased (1mA) until the experimenter observed visible displacement of fingers/hand, or the participant reported a maximum comfortable intensity. At this stage, we recorded (via our cameras) the participants' starting hand pose (relaxed) and final hand pose (end of stimulation). The difference between these two hand poses was then manually compared and annotated to *extract which fingers moved.* We opted for camera image annotations to avoid any noise from tracking systems [60, 80] contaminating our data. Thus, per trial, we labeled each of the participants' fingers as *flexion, extension, abduction, adduction,* or *no movement.*

### 4.4 Results

Overall, we found that all participants were successfully actuated from the stimulation points on the palm. Per participant, 22 ± 5 unique actuated finger poses were observed, and 49 ± 15% of electrode positions resulted in actuated finger poses. These results validated the feasibility of our approach—actuating from the contact points with the objects.

#### *4.4.1 Electrode positions*

Figure 9 shows, across all 378 stimulated electrode pairs on the palm, the electrode positions that resulted in visible involuntary movements. We observed that 49 ± 15% of electrode positions can be used to actuate fingers, ranging from 37% (P4) to 82% (P6). This is a striking result for EMS research, as palmar actuation is an unexplored target for EMS systems.

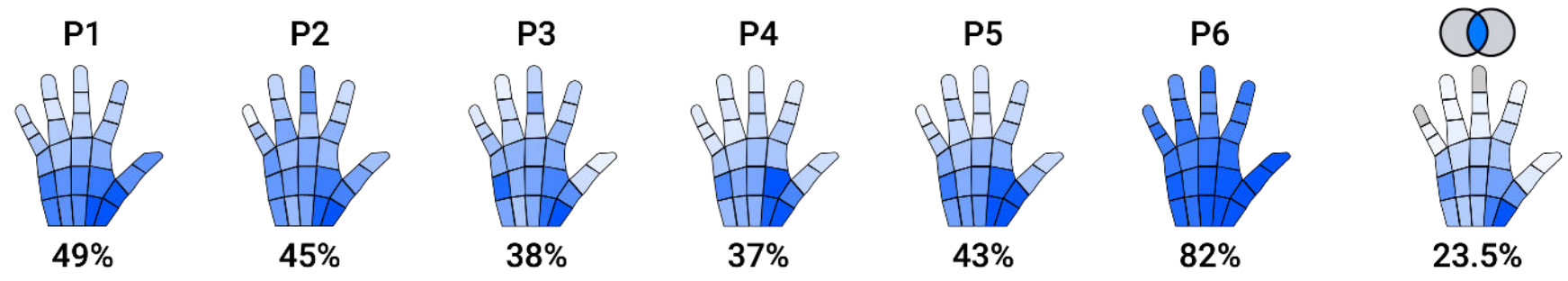

Figure 9: Palmar electrode positions that actuated the fingers, with EMS for each participant, and the intersection of all datasets. The color gradient depicts how often a location is involved in an electrode pair that resulted in a visible involuntary movement.

Finally, for the sake of completeness, we further intersect the observables (i.e., valid electrode pairs) to measure if there is any initial measurement of agreement on what electrode locations worked across participants. Despite the well-documented fact that EMS is very heavily participant dependent (e.g., [15, 42, 82, 90]), we found that 23.5% of these locations were shared across all participants. We believe this might inform useful hotspots for objects that aim at actuating fingers from the palmar side of the hand.

*4.4.2 Actuated poses*

Next, we turn our analysis to the main question: *what finger movements are possible from palmar-side EMS actuations?* We annotated each finger with four degrees of freedom (*flexion, extension, abduction, adduction*), yielding a total of 20 degrees of freedom and 3,124 different combinations of actuated movements. We found participants' hand could be actuated, from electrodes attached to the palmar side, to cause an average of 22 ± 5 different poses, with the most common observed poses being: *thumb abduction, thumb flexion, pinky flexion, thumb + pinky flexion* (these movements were observed for all six participants) and less common *index extension, middle + ring flexion*, etc. (these movements were observed only for two participants) Figure 10 shows the actuated poses that were observed for at least half of the participants, along with the exact numbers of participants for each actuated pose. The complete dataset is available in *Supplementary Material.*

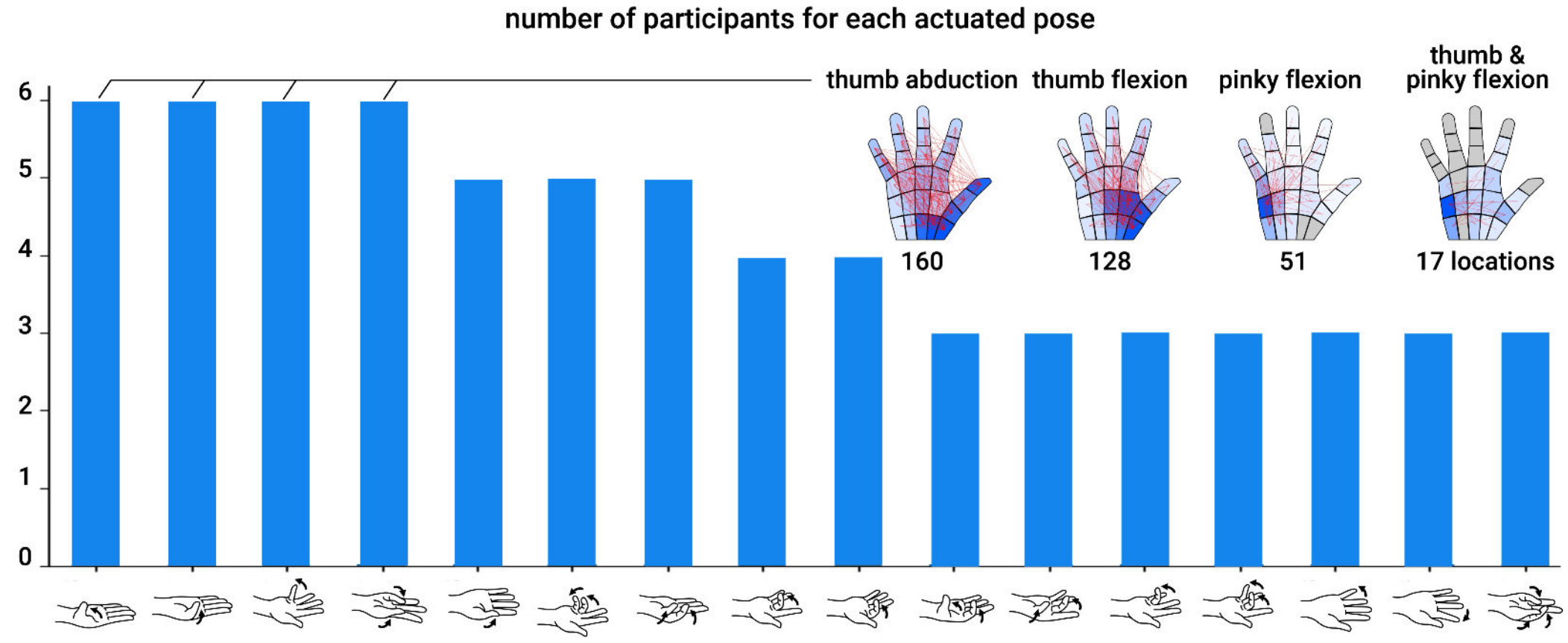

Figure 10: Participant count per actuated pose (zoomed to 50% of the participants; full plot in Supplementary Material). From left to right poses are: thumb abduction, thumb flexion, pinky flexion, thumb + pinky flexion, thumb adduction, ring finger + pinky flexion, thumb + index finger flexion, middle finger flexion, index finger flexion, thumb abduction + index finger flexion, thumb + middle finger flexion, ring finger flexion, middle finger + pinky flexion, pinky abduction, index finger abduction, and thumb + index finger + pinky flexion. The callouts show the union of electrode positions for each actuated pose observed on all participants' hands. Normalized color gradient depicts how often a location is involved in an electrode pair (red line) that resulted in a visible movement. Normalization is per actuated pose.

To illustrate the general pattern of electrode positions of the actuated poses, we show the union of electrode positions across six participants for the actuated poses that were observed for all participants. Callouts in Figure 10 detail the electrode positions (segments color-coded with a blue gradient) and pairs (two segments connected via a red line) that resulted in visible movements. The color gradient is normalized per participant for each actuated pose. We observed four distinct actuated poses that occurred reliably on all participants' hands—presented in the order of the number of electrode pairs that resulted in these observed poses: (1) **thumb abduction** (union: 160 different electrode pairs); (2) **thumb flexion** (union: 128 pairs); (3) **pinky flexion** (union: 51 pairs); and (4) **thumb + pinky flexion**, akin to a thumb-to-pinky opposition (union: 17 pairs).

For each participant, there were 59.3 ± 44.5 different pairs of electrode positions to actuate thumb abduction; 39 ± 24.6 positions for thumb flexion; 18.8 ± 5.5 positions for pinky flexion, and 3.8 ± 1.7 positions for thumb + pinky flexion. The fact that many of these poses can be achieved via a large number of locations is an important result in our concept, i.e., contact areas of the hand available during object manipulation and grasp can be leveraged by EMS systems to actuate the user's fingers.

Of these four observed EMS finger poses, three were *independent* poses (i.e., no other fingers were observed to move during actuation), specifically: (1) thumb flexion; (2) thumb abduction; and (3) pinky flexion; while the remaining was a *multi-finger* pose, specifically (4) thumb-to-pinky pinch. Moreover, two of the movements (i.e., thumb abduction and pinky flexion) were consistent across participants in both the poses themselves and the sets of electrode positions that reliably worked: two pairs of electrode positions could actuate thumb abduction for all participants, and three electrode pairs for pinky flexion. While the goal of our empirical validation was not to search for universal electrode positions, the preliminary agreement shown in the intersection of the electrode positions provides starting points for calibration.

## 5 PROOF OF CONCEPT: INSTANTIATING A SYSTEM TO IMPLEMENT OBJECT EMS

To help readers replicate our design, we provide the necessary details. As shown in Figure 11, for an object to implement *ObjectEMS,* it must realize three components: (1) a stimulation module, (2) a sensing module, and (3) a control module to close the loop between sensing and stimulation.

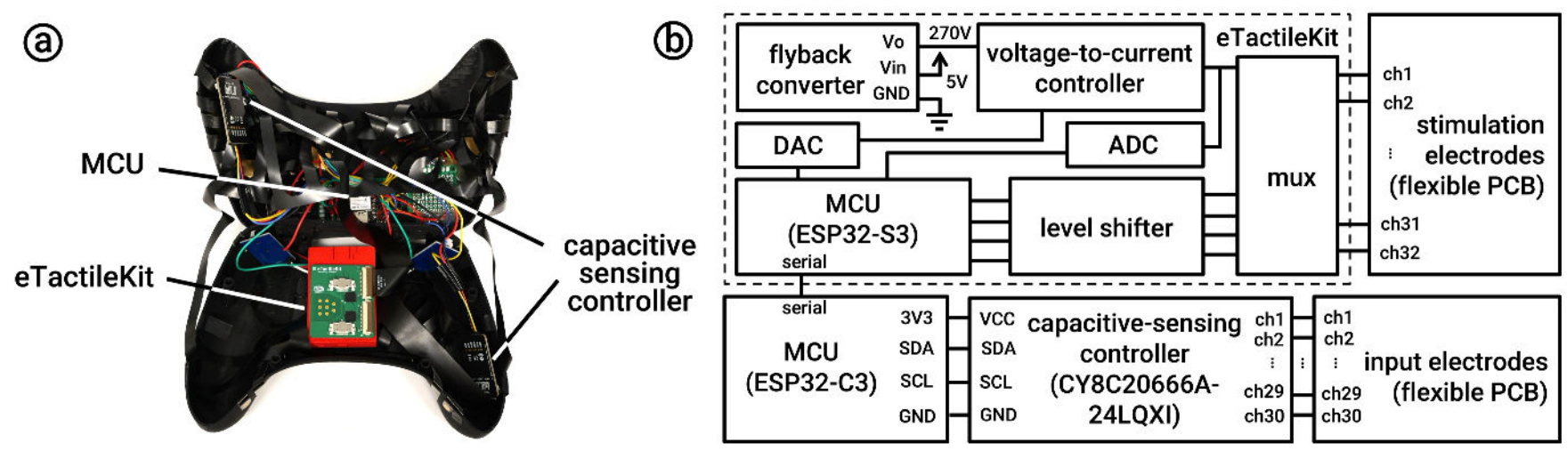

Figure 11: (a) High-level modules used to implement our system in the case of the game controller, and (b) their schematic.

### 5.1 Stimulation module

The stimulation module requires a compact, yet high-channel EMS stimulator, suitable for being embedded in hand-held objects (e.g., a game controller). For this, we selected the *eTactileKit* open-source stimulator [67], which satisfies both requirements with 10 mA (sufficient for finger actuation).

### 5.2 Sensing module

Realizing hand-pose detection on objects via capacitive touch requires a high-channel capacitive sensor at a small footprint. For this, we selected the *CY8C20666A-24LQXI* from *Infineon* (breakout from *Bela* [1]) with 30 sensing channels and communicates via I2C to our microcontroller.

### 5.3 Control module

#### *5.3.1 Information flow of the control module*

Figure 12 shows our control approach in *ObjectEMS.* Applications (such as the video game in Figure 1) send requests over serial with a list of actuated poses (e.g., a thumb flexion for power-up). This triggers the control to check if the user's current hand-pose is conducive to stimulation (which we will detail later). If the hand-pose is available for the requested actuation, our system activates the electrodes corresponding to the requested actuation (this mapping is fixed on the controller). During stimulation, the system constantly monitors the pose contact via capacitive sensing (of capacitive electrodes associated with the stimulated electrodes) and halts the stimulation whenever the capacitive sensing value is below the threshold (e.g., the hand leaves the surface, etc.). This is not the only technical way one can implement such features, e.g., impedance measurements have also been used for a similar purpose [66].

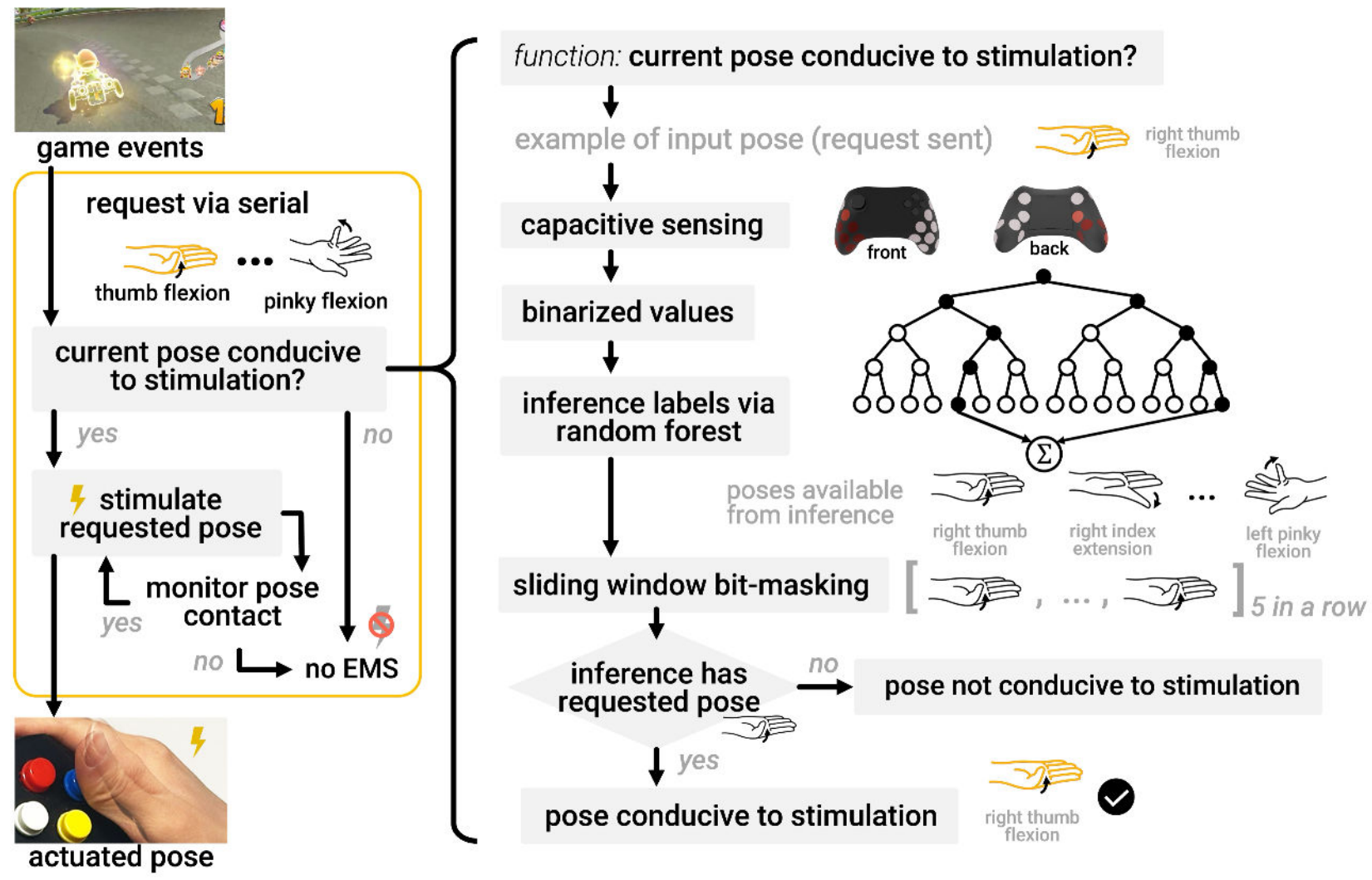

Figure 12: Information flow of our control module and the algorithm inferring poses conducive to stimulation.

### *5.3.2 Inferring if a pose is conducive to stimulation*

The right side of Figure 12 depicts how our system checks if the current pose is conducive to stimulation by classifying the current pose and comparing it against a list of accepted poses. Inspired by [8], once a complete capacitive array is read via I2C (~50 Hz in our current firmware), its values are binarized. This binarized capacitive "image" of the current pose is then used for multi-label predictions of possible actuated poses via a Random Forest classifier [48]—again, it is important to highlight that our main contribution is not innovate in multi-touch sensing nor to exhaustively compare our capacitive sensing alternatives (see decades of work on these topics alone in our field, in *Related Work*); instead, this implementation provides the required sensing to detect the user's poses and trigger electrical muscle stimulation from *electrodes placed on the surface of objects* that the user interacts with—this novel strategy is our key contribution. Moreover, to prevent any "flickering" from our simple classifier (where predictions can unstably jump between different classes), we added a sliding window that prevents the algorithm from continuing until five consecutive predictions of the same class are observed. Finally, the inferred pose is checked against the list of possible poses for a given application and *ObjectEMS* device, which is saved in the application's code. If the result is considered a valid pose for this device/application, the stimulation is allowed to start.

### *5.3.3 Classification accuracy (technical experiment)*

To evaluate the performance of our Random Forest classifier for capacitive pose sensing with our game controller, we recruited 10 participants (six female, four male; 25.6 ± 2.8 years old; all right-handed) to collect pose sensing data. This technical validation was approved by our ethics review board (IRB19-1949). In this experiment, we asked participants to pose with the controller in five ways given only verbal instructions (e.g., "hold the controller with both hands to move its joystick and press its buttons", "hold the controller with both hands but do not use the joystick nor buttons", "hold the controller with your left hand", "hold the controller with your right hand", and "hold the controller in any way you like that you do not consider holding as if you are playing"). After each sample was collected (capacitive readouts for all channels), we asked them to put down the controller and move to the next pose, repeating this process 10 times for each pose. We conducted two evaluations to determine our classifier's performance with both known and unknown (i.e., unseen) user data, grounded in standard methodology. First, we performed a stratified 5-fold cross-validation across all participant data. For each participant, poses were partitioned into five folds: Four folds (8 samples per pose per participant, for a total of 400 samples) were for training, and the remaining fold (2 samples per pose per participant, for a total of 100 samples) was for evaluation. Across all iterations, our model achieved >90% accuracy, meaning that our classifier is highly robust at detecting poses for calibrated users (i.e., their data was included in the training set). Second, we conducted a Leave-One-Participant-Out (LOPO) cross-validation where nine participants' data was used for training and the remaining one for evaluation (i.e., the unseen participant) across 10 iterations. As shown in Figure 13 (b), our classifier achieved 74.10% (SD=11.42%) accuracy, with the "hold with left hand" and "hold with right hand" achieving >90% classification accuracy, as depicted in Figure 13 (a). This confirms that our model is still able to detect new unseen subjects with some degree of confidence (i.e., their data was not part of the training set). It is important to note that while this recognition accuracy was deemed sufficient for our demonstrative applications, there is ample work in this area (see *Related Work*) showcasing more advanced approaches that can dramatically enhance accuracy bounds and handle complex edge cases.

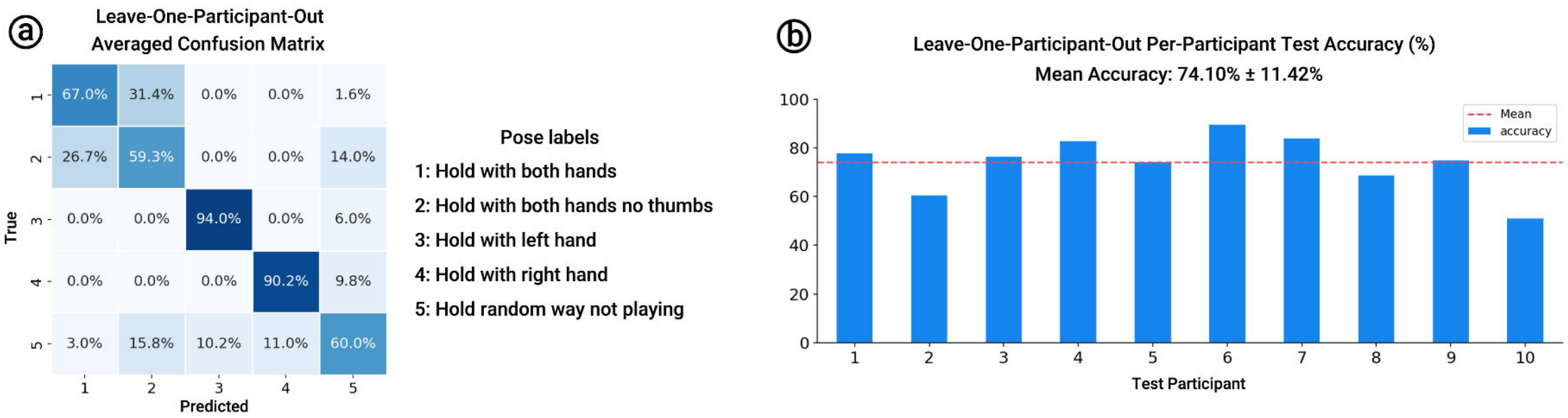

Figure 13: (a) Leave-one-participant-out averaged confusion matrix. (b) Leave-one-participant-out per-participant test accuracy.

### 5.4 Fabricating electrodes

Both the sensing modules and the stimulator modules are connected to a set of electrodes. There are a plethora of electrode types (each derived from a unique manufacturing process) can be used to implement *ObjectEMS*, as depicted in Figure 14: (a) pre-gelled adhesive electrodes, (b) dry carbon electrodes, (c) dry 3D printed electrodes (conductive filament), or (d) dry electrodes made from PCBs (typically flexible but rigid are possible, too). While pre-gelled electrodes are most common in EMS systems, dry electrodes are popular in electrotactile (e.g., [24, 81, 89]) and are also used in commercial EMS products like *Teslasuit* [87]. We selected dry electrodes in all our exemplary implementations to make objects easy to detach.

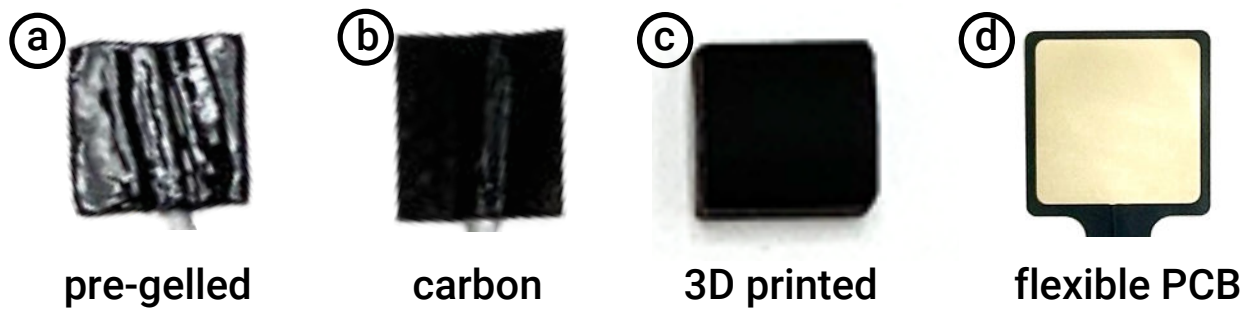

Figure 14: Different electrodes: (a) pre-gelled adhesive; (b) carbon; (c) 3D printed with conductive filament (Proto-pasta Conductive PLA); (d) flexible PCB made from copper with ENIG (electroless nickel immersion gold) treatment.

### 5.5 Combining electrodes

Since both our sensing module (capacitive touch) and our output module (electrical muscle stimulation) function based on conductive electrodes, they can be combined using varied layouts. Figure 15 shows possible topologies when combining input and output electrodes (using the example of the most versatile, flexible PCB, electrode): (a) *side by side*: input and output electrodes adjacent to each other; (b) *sensing around stimulation*: the input electrode surrounds the an isolated output electrode (c) *stimulation around sensing*: the reverse of the previous topology; (d) *collocated*: input electrode is on the back of the same substrate (dual-sided PCB)—since capacitive touch does not require direct contact, this still works for sensing (reverse is not possible as stimulation requires skin contact); and, finally, (e) *single unified electrode:* input/output time-multiplexed.

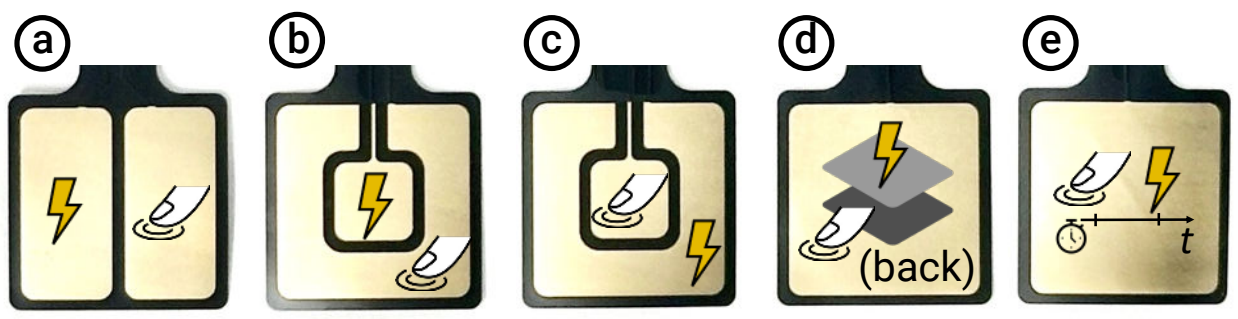

Figure 15: Layouts for input/output electrodes: (a) side by side; (b) sensing around stimulation; (c) stimulation around sensing; (d) collocated; and (e) time-multiplexed.

### 5.6 Making an object with ObjectEMS

Figure 16 shows different approaches for implementing *ObjectEMS* into an object: (a) akin to a do-it-yourself (DIY) approach, by attaching the stimulation and sensing electrodes to the surface of the object—this allows exploring sensing/stimulation layouts on the objects' geometry; (b) a "sleeve" (i.e., a rigid or flexible material packing multiple electrodes) enabling to reuse and apply *ObjectEMS* to different objects with similar shapes (e.g., different game controllers, or different bottles, as shown in *Applications*); and, (c) designing a custom object that directly embedded electrodes and stimulator inside the object (e.g., 3d printed).

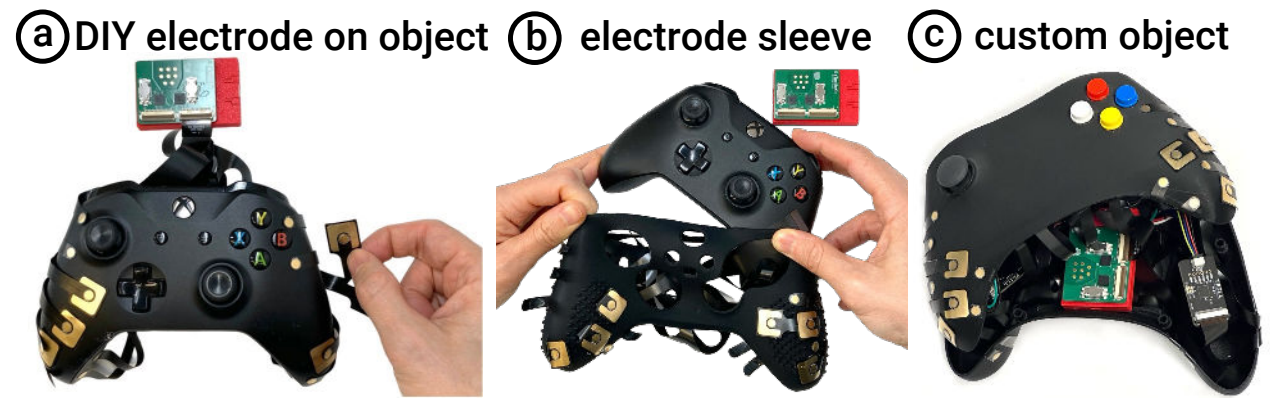

Figure 16: Different approaches for instrumenting an object.

## 6 USER STUDY: OBJECT EMS VS TRADITIONAL EMS

In our main user study, we compared *ObjectEMS* with the traditional approach (i.e., EMS electrodes worn on the body) as shown in *Walkthrough*. We asked the participants to first calibrate EMS by themselves and then experience one of our applications: playing Mario Kart with force-feedback via EMS using both *ObjectEMS* and the traditional approach. This allowed us to ask: (1) Does *ObjectEMS* make EMS calibration easier for the user? (e.g., they calibrate faster with *ObjectEMS* with fewer times of trial-and-error attempts); and (2) How does the user feel using an EMS application implemented by *ObjectEMS* vs. a traditional body-worn EMS while engaging in social interactions?

### 6.1 Study design

We followed the study design of Takahashi et al. [80], which was set up to observe users' reactions to a novel form factor of EMS compared with a traditional method. In this design, we invited participants to complete two tasks: in Task#1 we asked participants to calibrate the EMS by themselves, while in Task#2, they used an EMS application (force-feedback while gaming, same as in *Walkthrough*) while engaging in social interactions (e.g., opening a door to a person). Participants completed these tasks via *ObjectEMS* and traditional EMS (condition order counterbalanced).

#### *6.1.1 Interface conditions*

Participants experienced two EMS conditions: traditional **body-worn EMS** (i.e., the approach used in all interactive-EMS prior works [18]) and our ***ObjectEMS*** device. The only difference between conditions was the **location of the electrodes** (i.e., attached to the participant's body vs. attached to the object); all other factors were kept consistent between the two conditions (e.g., the participants used the same calibration interface and played the same *Mario Kart* game with the same game controller).

#### *6.1.2 Task#1: calibrate EMS*

Participants were tasked to calibrate three gestures: left thumb abduction to move the joystick, left pinky flexion, and right pinky flexion, together making the rumble effect. In the body-worn EMS condition, participants received an image showing the anatomical structure of muscles on the forearm and hand as a reference; in the *ObjectEMS* condition, the game controller was prepared with electrodes at locations based on our findings from *Empirical Validation*. In both conditions, participants had to decide and adjust electrode locations and fine-tune the stimulation *all by themselves*. We let the participants calibrate either EMS condition for up to 20 minutes.

#### 6.1.3 Task#2: using EMS

In this task, we instructed participants to play a *Mario Kart* game (similar to *Walkthrough*) for four minutes and experience two types of haptic effects based on the three calibrated gestures: steering the joystick (left thumb abduction) when the participant hits banana peels and the rumble effect (alternating between left pinky flexion and right pinky flexion) when the participant drives on lava rocks. The haptic effects were scripted in the game so that every participant experienced them three times for both effects. To invite participants to think about their use of EMS in a wider context (i.e., that involves the perception of others as well), we included pre-scripted social interactions—these were also borrowed from the study design of Takahashi et al. [80], which included similar social interactions. As such, while playing the *Mario Kart* EMS game, participants were asked to open the door to let a person into the study room (a second experimenter knocking at the door), to pick up items that fell from the study table (purposely done by the primary experimenter), and to high-five with the primary experimenter right after they completed the game. After the first two interactions, participants were asked to resume the game, including recalibrating or reattaching any hardware if necessary. Finally, as in [80], these simple social interactions were added to elicit qualitative feedback on whether EMS conditions affect interactions beyond the in-game EMS haptic effects. All interactions were timed equally across both conditions (e.g., door knocking happened after participants experienced two haptic effects, and the objects fell down from the table after the fourth haptic effect).

### 6.2 Apparatus

In both conditions, electrical stimulation was delivered via a medically-compliant stimulator (*HASOMED*). We used dry electrodes in the *ObjectEMS* condition (i.e., attached to the object) and used pre-gelled electrodes in the traditional body-worn EMS condition. We opted for pre-gelled electrodes for the body-worn EMS condition so that participants could easily attach the electrodes to their bodies. Note that this choice favored the body-worn EMS approach against our condition, as dry electrodes cause more tingling sensations on the skin.

### 6.3 Participants

We recruited 12 participants (five female, seven male; 26.1 ± 2.1 years old; all right-handed), compensated monetarily for their time. This study was approved by our ethics review board (IRB19-1949).

### 6.4 Task#1 Results

Figure 17 shows the quantitative results. We found participants spent significantly less time completing the calibration for *ObjectEMS* (M = 7.1 mins, SD = 2.9) compared to body-worn EMS (M = 13.5 mins, SD = 0.2). On average, participants made 3.83 ± 4.75 electrode location adjustments on *ObjectEMS*, and 9.67 ± 9.7 adjustments on body-worn EMS. One of the participants did not complete the calibration within our 20-minute timeframe; in this case, this participant's completion time was set to the maximum (20 minutes) for analysis. Next, we turn our attention to participants' comments.

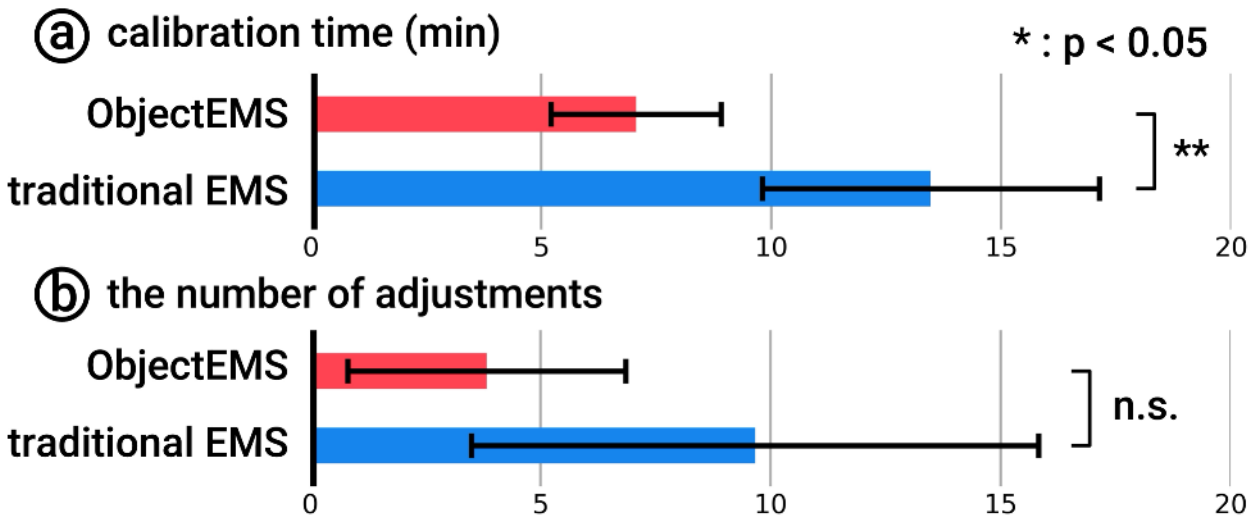

Figure 17: Results for (a) calibration time, (b) number of electrode location adjustments. (Error bars depict 95% confidence intervals).

#### 6.4.1 Advantages of strategic electrode locations on object.

We noted many statements where participants detailed that they found it "easier" (P3, P4, P5, P6, P10) or "more straightforward" (P9) to calibrate *ObjectEMS*, even if they still needed to adjust the electrode locations on the game controller. For example, P10 elaborated that "[electrodes] are (…) quite close to what I want" in the *ObjectEMS* condition. In contrast, "the search space is very large" (P11) or "required me to have a little bit more awareness of the physiology (...) to know where to mount electrodes to get the right effect" (P5) in the body-worn electrode condition.

*6.4.2 Challenges of ObjectEMS.*

Although most participants found it easier to calibrate *ObjectEMS*, two (out of 12) participants also mentioned difficulties. P8 noted an added challenge of "where [electrodes] map onto the controller", and P12 noted, "depends on the size of your hand".

### 6.5 Task#2 Results

To prime participants to reflect on their use of EMS in a wider context, our study required participants to engage in social interactions (e.g., answering a door for someone) while using an EMS application, which allowed us to elicit positive feedback but also identify some challenges regarding *ObjectEMS*.

*6.5.1 Benefits of EMS without electrodes (on the user).*

Participants praised the retained mobility of not having to attach anything to their bodies. They found it was easy to put down and pick up to use *ObjectEMS* (8 out of 12 participants), while mobility was impaired by cable length in body-worn EMS (9 out of 12). For example, P7 elaborated, "It was much easier to just set down the controller and go to the door rather than peel everything off," while with body-worn EMS, "you're almost chained down a little bit because they [cables] only reach so far." (P7) Participants also mentioned that there was less friction of getting back to the EMS application with *ObjectEMS*: "my muscle memory of holding the controller was much easier and easy to recall, compared to putting the electrodes [back] on the body" (P6) and "I was not worried about setting down the controller [with *ObjectEMS*] because the controller had already been set down with initial calibrations anyway." (P5) Figure 18 shows an example of how (a) a participant tried to avoid taking off all of the electrodes by opening the door with a challenging posture, compared to (b), participant could simply put the controller down and walk to the door. Figure 19 shows a participant picking up a fallen item. In the traditional EMS condition, the participant had to pull the cable to ensure it was long enough for them to reach the item, whereas in *ObjectEMS* condition, they could simply put down the controller to pick up an item on the ground.

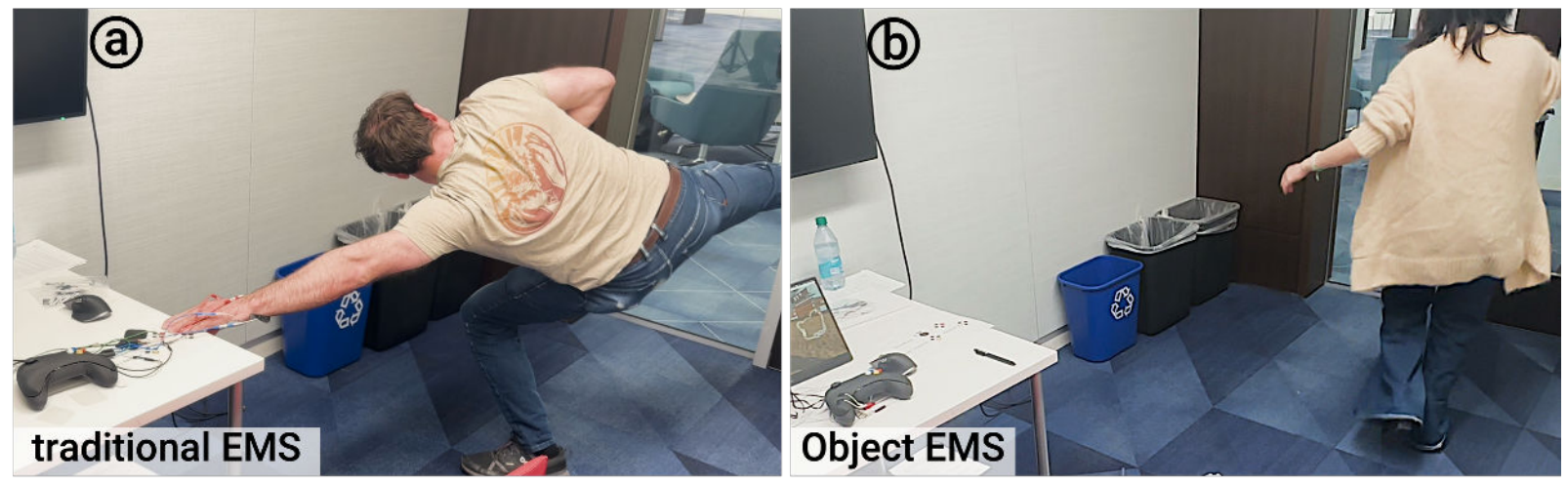

Figure 18: Participants opened the door in (a) traditional body-worn EMS condition, where they were tethered by the cables, and (b) *ObjectEMS* condition, where they could easily put down the controller.

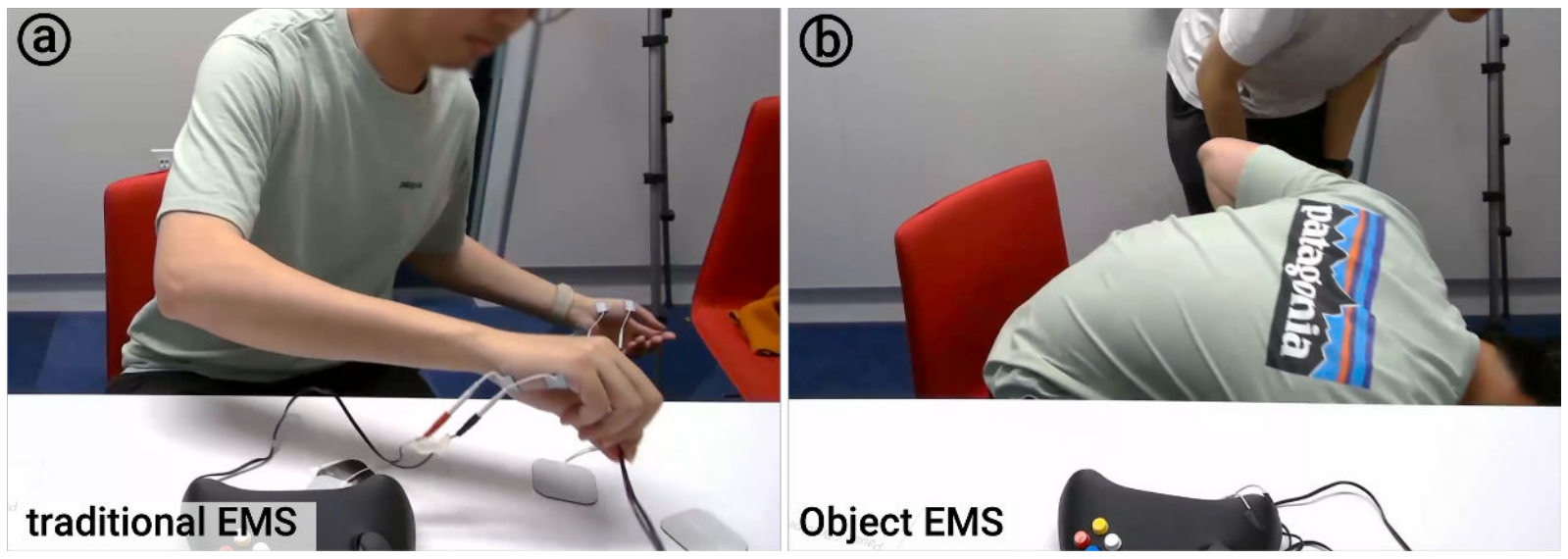

Figure 19: A participant picked up an item on the ground in (a) traditional EMS condition, where they had to ensure the cable length could reach the ground, and (b) *ObjectEMS* condition, where they could simply reach the ground without being limited by the cable.

*6.5.2 Challenges of ObjectEMS*

As expected, limitations were also voiced, such as *ObjectEMS* felt less intense in terms of force (3 out of 12) and required stable contact with electrodes (8 out of 12). For example, P2 mentioned, "the only difficulty was that when I came back to the controller, I had to make sure I was grabbing it in the exact same way", and that "making sure that my thumb is making contact with the [electrode] pad caused a little bit of discomfort" (P8).

#### *6.5.3 Safety*

While one of the participants brought up that "I probably would be kind of worried about touching it [the objects] at all." (P9) when imaging daily objects were instrumented with EMS, three of the participants stated that *ObjectEMS* made them feel safer to use compared to body-worn EMS, because, for instance, "if it's on object, if I feel uncomfortable, or I think it's dangerous, I can just throw or drop the object" (P12).

## 7 APPLICATIONS

Finally, we demonstrate a set of EMS interactions where users do not need to carry any electrodes around their bodies. This is unprecedented in EMS, as all prior works require body-worn electrodes. These applications were designed to illustrate different ways that future researchers can explore *ObjectEMS*, depending on the type of manufacturing process used, from rapid prototyping via flexible electrodes, sleeves with electrodes that wrap around objects, to designing objects (from scratch) featuring built-in electrodes.

### 7.1 Application #1: a "no-strings attached" EMS guitar

Our first application is depicted in Figure 20, which illustrates a user who attached an *ObjectEMS* set of electrodes (and an accompanying external enclosure with the hardware) to the back of their guitar's neck. Actuating from this vantage point allows the user to feel force feedback on their thumb to press harder on the string (via the thumb flexion gesture found in our *Empirical Validation*). We contrast how, unlike in previous explorations of EMS for playing guitar [60, 79], with *ObjectEMS* the user's hand is completely free from many tethers and electrodes that plague EMS systems.

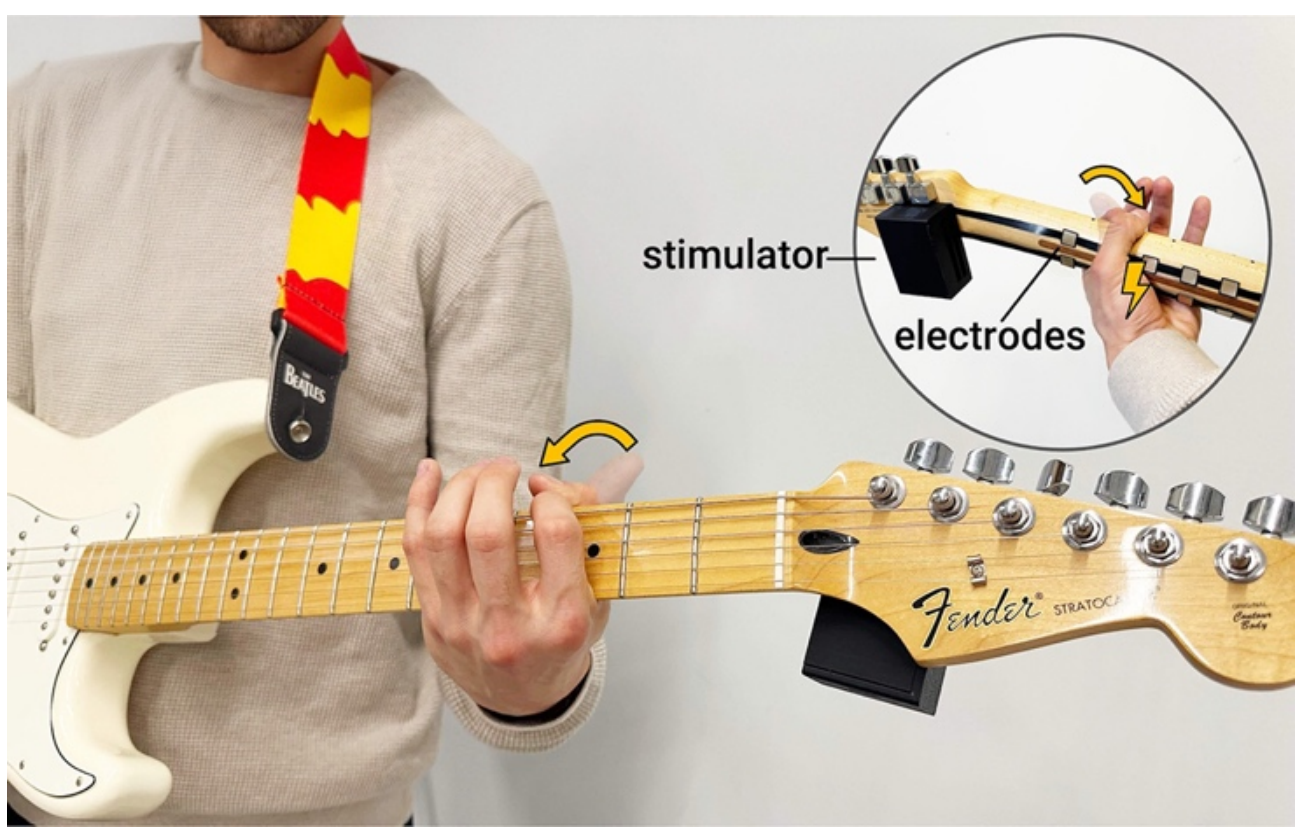

Figure 20: By attaching a stimulator and electrodes to a guitar, ObjectEMS assists users with playing the guitar by flexing their thumb to add force to the press-down gesture.

### 7.2 Application #2: actuating objects via *ObjectEMS* sleeves

In our second application, depicted in Figure 21, we demonstrate how *ObjectEMS* can also be applied by using electrode "sleeves", enabling rapid prototyping with actuated objects. In Figure 21 (a), a 3D-printed cylindrical "sleeve" holds a water bottle prototype that, via *ObjectEMS*, reminds its user to hydrate if their sips are too shallow. It achieves this using a shake gesture, shown in Figure 21 (b), which is performed by alternating pinky flexion and thumb flexion. As shown in Figure 21 (c), the same sleeve can be reloaded with another prototype object, such as a spray can that (d) can remind its users to shake prior to use—yet, unlike prior work, it does not require body-worn electrodes [28, 54].

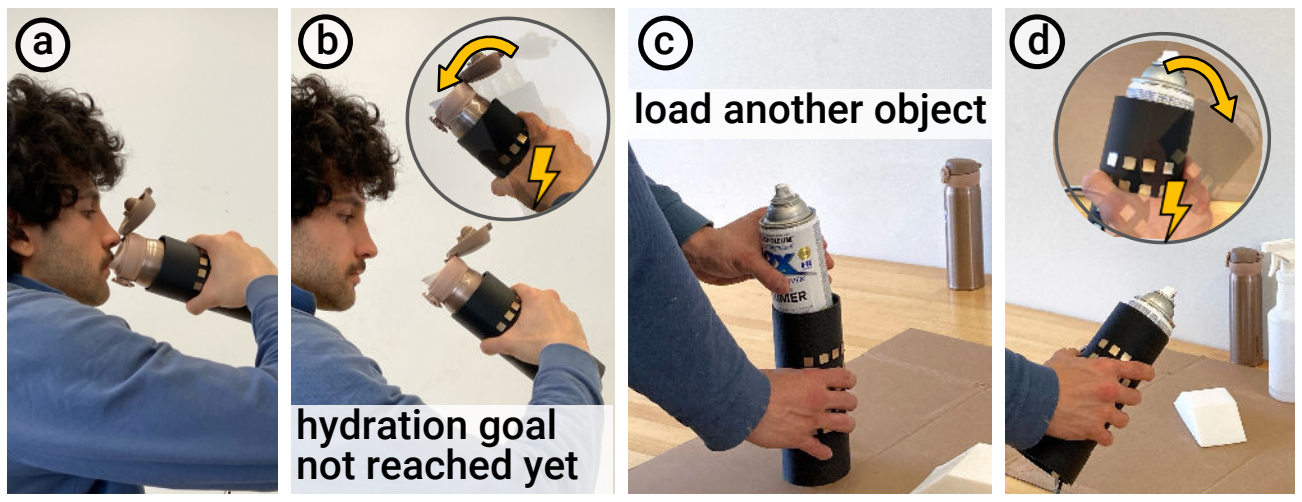

Figure 21: *ObjectEMS* can also be deployed using electrode "sleeves" that enable quick testing of different actuated objects, such as this (a-b) water bottle, or (c-d) a spray can.

### 7.3 Application #3: self-contained (EMS) actuated tangibles

Finally, our third application, depicted in Figure 22, illustrates an entirely 3D-printed object that features *ObjectEMS* inside. This exercise dumbbell includes all the sensing and stimulation hardware, as well as electrodes in the handle—to illustrate the integrated nature of this example, the electrodes were 3D printed on the handle using conductive filament and soldered to the stimulator using copper tape and wires. As a result, the dumbbell becomes interactive: (b) it uses thumb abduction to pace the user's biceps curls; and (c) index extension to indicate when the set is over, and a rest period starts.

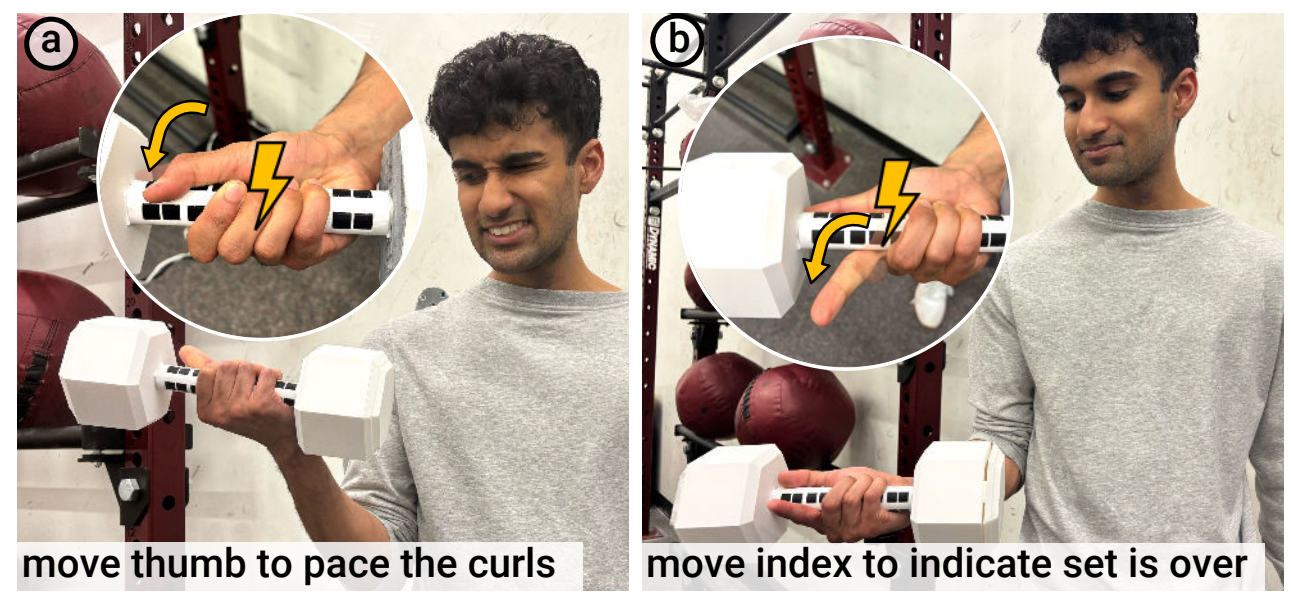

Figure 22: A 3D-printed dumbbell featuring all of *ObjectEMS* hardware and conductive filament-based electrodes.

## 8 LIMITATIONS & FUTURE WORK

As the first exploration in embedding electrodes in objects that user interact with (rather than in the body), ours is not without its limitations, which we believe are fertile ground for future work: (1) naturally, this technique is only suited for body areas that make bare contact with objects, which is why our investigation and two studies solely focused on the hands (the body surface that most commonly makes contact with objects)—while there might be other possible situations (e.g., neck while resting on furniture or exposed ankles) these tend to be less common, and, even hands can be unavailable (gloves); (2) instrumenting electrodes on objects is not suitable for all kinds of objects, such as if the objects must be passive (e.g., disposable, sustainable, etc.) or even weather proof; (3) while EMS is likely the most miniaturized form of force-feedback, embedding electrodes, stimulators, and batteries requires a minimum object size that respects the necessary footprint of electrodes and hardware—this limits using this technique inside of small objects (e.g., chopsticks), but still leaves open the option of attaching electrodes plus an extra external enclosure for the hardware (as intentionally depicted in some of our applications); moreover, as a technique based on electrical muscle stimulation, it inherits all of EMS' well-documented limitations, namely: (4) it requires calibration (as shown in *User Study*, participants needed to adjust electrode positions); (5) it creates an unwanted tactile sensations ("tingling") on the skin (this is slightly exacerbated with dry electrodes)—while our technique is electrode-agnostic (gelled electrodes can also be used), dry electrodes are better suited to avoid altering the existing friction of objects; (6) as usual with EMS, it is possible that different body poses slightly modulate the resulting actuations—while our study was extremely exhaustive (2,268 trials in total) we did not explore a wide range of body poses; and, (7) with EMS, differences between individuals are common (hence the need for calibration) [15, 42, 82, 90]. While this is an active area of EMS research, it is worth highlighting that we did surprisingly find preliminary evidence for locations in the hand that agreed, even across subjects.

## 9 CONCLUSION

We proposed a new route for electrical muscle stimulation (EMS) that does not rely on attaching electrodes to users, but instead embeds electrodes on objects that users might interact with. In an empirical validation focused on electrically stimulating the palmar side of the hand (i.e., locations that would normally be involved in grasping), we found that it is possible to move the fingers in four different poses via EMS. To demonstrate this technique, we developed self-contained proof-of-concept objects with embedded electrodes and a stimulator inside the object, such as game controllers, guitars, or dumbbells—all with built-in force feedback via on-object EMS. We contrasted our approach with traditional EMS in our user study and observed that participants found it easier to calibrate *ObjectEMS*, and they spent less time calibrating than in traditional EMS. They also found it easy to switch between *ObjectEMS* and other situations (e.g., social interactions); in comparison, they felt their mobility was impaired by the cable length of the traditional EMS approach.

We believe that this fresh perspective on EMS, which frees up the user's body from the burden of electrodes, enables a new pathway to explore many of the interactive domains where EMS research has been focused, but has lagged due to its dependency on body-worn electrodes, such as casual or rapid interactions with tools and everyday objects.